\documentclass[10pt,letterpaper]{article}
\usepackage{opex3}
\usepackage{verbatim}       

\begin{document}

\title{Characterization of an InGaAs/InP~single-photon~detector at 200~MHz~gate~rate}

\author{Chris~Healey, Itzel~Lucio-Martinez, Michael~R.~E.~Lamont, Xiaofan~Mo and Wolfgang~Tittel}

\address{Institute for Quantum Information Science, and Department of Physics and
Astronomy, University of Calgary, 2500 University Drive NW,
Calgary, Alberta, Canada, T2N 1N4}

\email{chealey@qis.ucalgary.ca} 



\begin{abstract*}
We characterize a near-infrared single-photon detector based on an InGaAs/InP avalanche photodiode and the self-differencing post-processing technique.  It operates at gate rates of 200~MHz and higher. The compact, integrated design employs printed circuit boards and features a semiconductor-based self-differencing subtraction implemented with a fully differential amplifier.  At a single-photon detection efficiency of $6.4\%$, the detector has a dark count probability of $9$x$10^{-7}$ per gate, an afterpulse probability of $6.3\%$ per detection event, a detection time jitter of $\sim$150~ps, and a  dead time of  5~ns (equivalent to one gate period). Furthermore, it can be operated as a standard photodiode, which benefits applications that require detecting single photons as well as strong light signals.\\ \\

\end{abstract*}







\section{Introduction}
\label{sec:intro}  

The need for high gate-rate single-photon detection in the near-infrared is well established~(see \cite{GisinQuantumCryptography,InoueSineWaveGating,ShieldsHighSpeedSPDNearInfrared,GenevaPracticalFastGateRate} and references therein). Avalanche photodiode (APD)-based single-photon detectors (SPDs) are highly suitable for these tasks because of their ease of operation, and have been commercialized by, e.g., idQuantique and Princeton Lightwave Inc.
Initially, an \textit{afterpulsing} effect  limited the use of these detectors to gate rates of a few MHz~\cite{RibordyPerformanceOfAPDs}. This impediment has been removed recently by means of clever electronic post-processing ~\cite{InoueSineWaveGating,ShieldsHighSpeedSPDNearInfrared,GenevaPracticalFastGateRate}. However, due to the sophisticated nature of these approaches, high-rate detectors have not yet been widely adopted. Here we give, for the first time, a detailed description of a high-rate SPD based on the self-differencing approach \cite{ShieldsHighSpeedSPDNearInfrared,GenevaPracticalFastGateRate}. Our detector comprises key-components on printed circuit boards, including a semiconductor-based subtraction, and features characteristics typical for high-rate single-photon detectors. Furthermore, when not being used in the photon counting regime, it operates as a standard photodiode. The possibility of rapidly switching between the two different operation modes is interesting in view of applications that require detecting single photons as well as strong light pulses, e.g. the quantum key distribution system described in~\cite{ItzelQC2QKD}.

\begin{figure}[tbh]
\centering
\includegraphics[width=12cm]{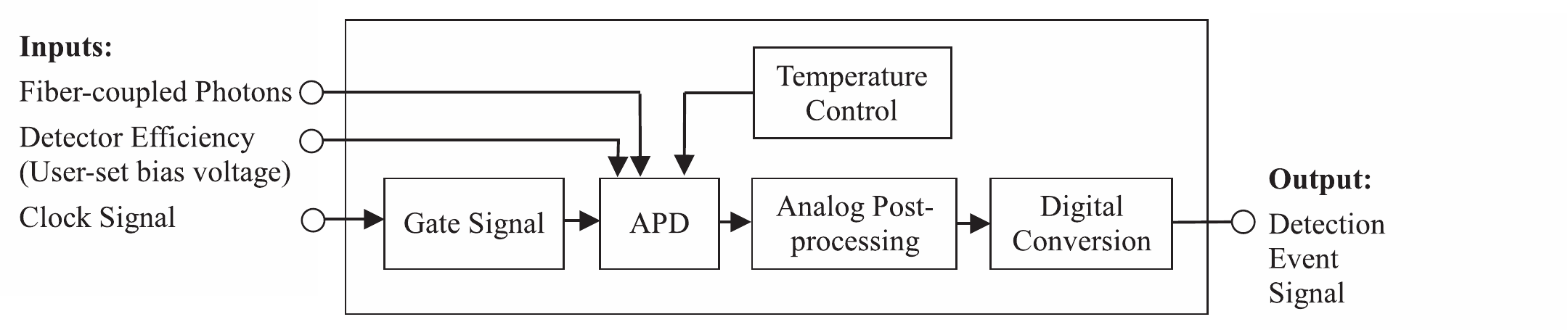}
  \caption{A functional block diagram for schemes utilizing an avalanche photodiode for single photon detection.}
\label{apdFunctionalDiagram}
\end{figure}

\section{High rate single-photon detection with APD's}
\label{sec:apds}  

\begin{figure}[t]
\centering
\includegraphics[width=10cm]{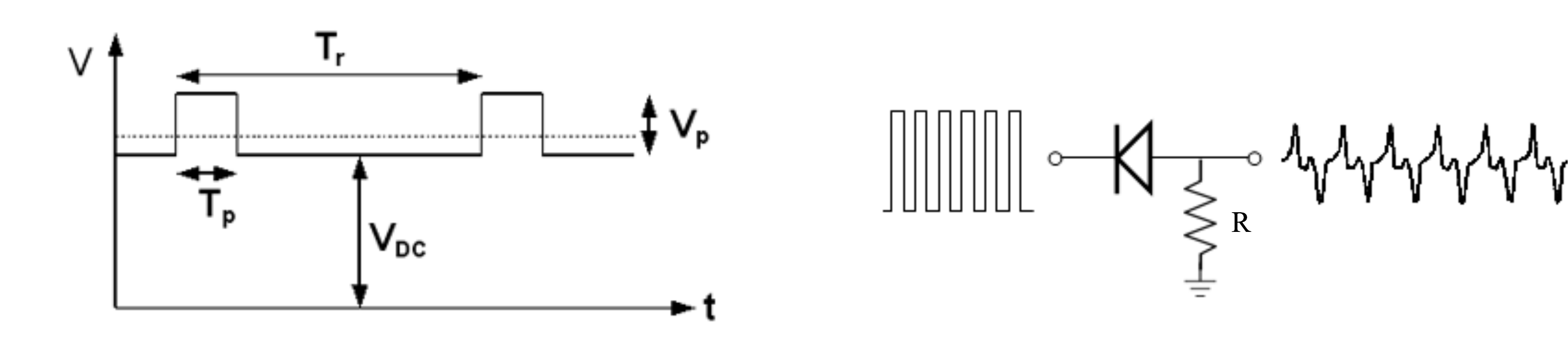}
  \caption{Typical operation of an APD in Geiger mode: a few-volt pulse V$_{p}$ is added to a reverse-bias of voltage V$_{DC}$, and applied to the cathode of the APD.  The pulse width T$_{p}$ is on the order of nanoseconds. A large deadtime of several $\mu$s is applied between subsequent gate pulses to limit afterpulsing effects. The dashed line indicates the breakdown voltage.}
\label{geigerGating}
\end{figure}

A functional diagram for using an APD in a typical near-infrared single-photon detection scheme is shown in Fig.~\ref{apdFunctionalDiagram}.  The APD is Geiger gated as illustrated in Fig.~\ref{geigerGating}).  A pulse of a few volts, V$_P$, nanosecond duration, T$_p$, and repeating at MHz rates, f$_g=1/$T$_r$, is added to a DC bias voltage, V$_{DC}$ (typically $\sim$60 V), and reverse-biases an APD beyond breakdown voltage. When a photon is incident on the absorption region of the APD substrate, it can photoionize a charge carrier.  This charge carrier undergoes a large acceleration due to the bias, and impact-ionizes other charge carriers to create an avalanching effect, eventually generating a macroscopic current creating a voltage across a current-limiting resistor. Increasing the excess voltage (i.e. the voltage above breakdown) can result in high non-linear gains and thus higher efficiencies, but also results in trapping more charge carriers in bandgap defects created by manufacturing constraints. These trapped charges can be released during subsequent gates creating spurious detection events, a phenomenon known as afterpulsing.  The APD is cooled to a temperature of $\sim-30^{o}$C to reduce dark counts; however, as the temperature is reduced, the lifetime of trapped charge carriers increases.  Cooling then creates a tradeoff between dark counts and afterpulsing (the ratio of dark count probability to efficiency reaches a minimum and is reasonably flat at temperatures below approximately $-30^{o}$C~\cite{WuSpikeCancellation}).  Current state-of-the-art commercial single photon detectors require dead times on the order of $\mu$s for trapped charge carriers to decay after a detection event, thereby limiting the clock rate and hence the detection rate.

\begin{figure}[b]
\centering
\includegraphics[width=10cm]{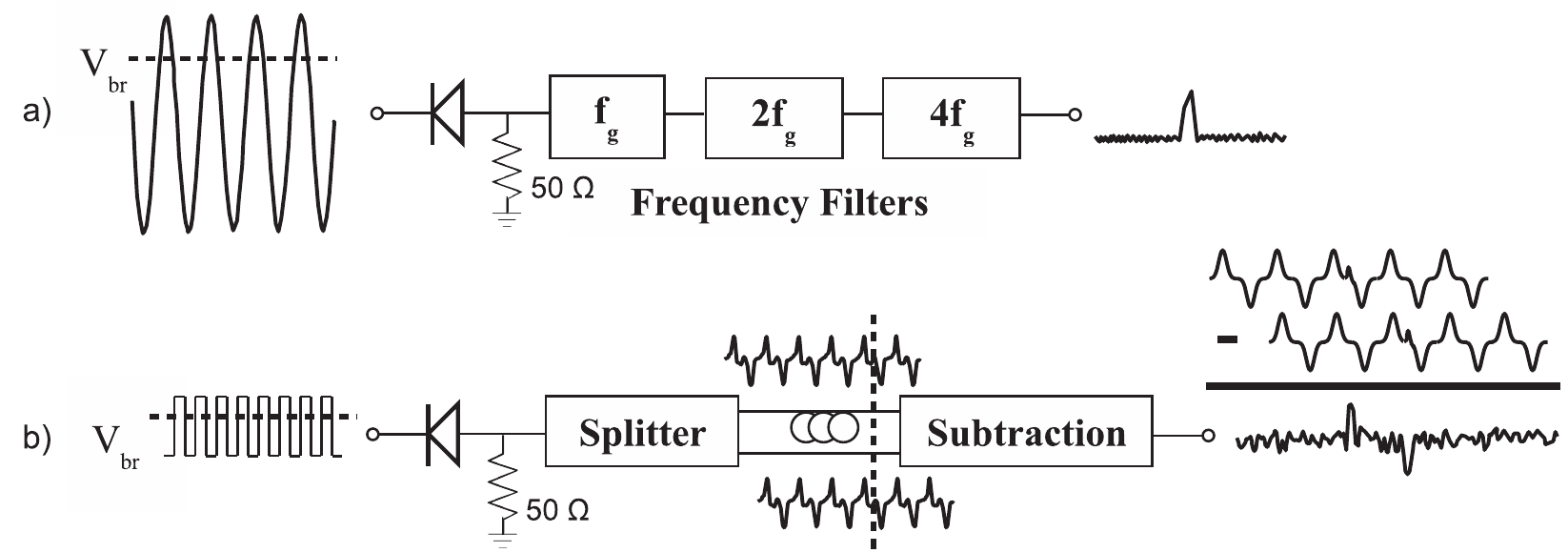}
  \caption{(a) When sine-wave gating an APD, the capacitive response can be filtered out to recover the avalanche signal. (b) The self-differencing technique: the APD output is sent into a splitter, one output  is delayed by one period, and the two outputs are subtracted to remove the now common-mode capacitive response, leaving only the signature of a photo-detection. For pedagogical reasons, the avalanche signal is exaggerated in size.}
\label{combinedSGandSDschematic}
\end{figure}

The APD responds to the gate pulse as a capacitor, and the charging and discharging creates a voltage on the output corresponding to the derivative of the gate pulse.  With low gate rates and nanosecond-wide gates, the avalanche signal can easily be discriminated against the capacitive response. As we increase the gate rate, this discrimination becomes more and more difficult: First, to reduce the required dead time, the amount of electrons generated per avalanche has to be lowered. This can be achieved by decreasing the gate voltage, leading to a smaller avalanche signal. Second, the gate width inevitably becomes shorter, the rise and fall times of the gating pulses become steeper, and the capacitive response thus increases in voltage. Hence, high gate rates, even if still compatible with ns-long gates, require sophisticated post-processing techniques to discriminate the avalanche masked by the large capacitive response.

One technique developed to do this uses a sine wave to gate the APD~\cite{InoueSineWaveGating}, as shown in Fig.~\ref{combinedSGandSDschematic}(a). Sine waves of 12 V are easy to generate at high rates; furthermore, the capacitive response of the APD is also a sine wave, so the output can be frequency-filtered with notch filters at the gating frequency and its first few harmonics to uncover detection events.  At high bandwidths, these analog filters become more and more difficult to design and build.

The so-called self-differencing technique (depicted in Fig.~\ref{combinedSGandSDschematic}(b)) takes the periodic output of the APD, gated using square pulses, and sends it into a signal splitter, thereby creating two identical (3 dB reduced) copies of the signal \cite{ShieldsHighSpeedSPDNearInfrared}.  One output is delayed by one period (or an integer number of periods~\cite{ShieldsMultiGHzSD}) and then the signals are subtracted. This removes most of the now common-mode capacitive response, leaving the processed avalanche signal sitting on a small background. A characteristic upwards peak followed by a downwards peak a period later is seen on the output (the polarity of these signals can change depending on which signal was subtracted from which).  We note that the maximum detection rate is limited to half the detector gate frequency, due to the necessity to have an adjacent gate without a detection event for the subtraction.

These techniques have been combined ~\cite{GenevaPracticalFastGateRate}; a sine-wave gating was used, and the APD output initially frequency-filtered to remove most of the common-mode capacitive response.  A self-differencing circuit was then used to increase the signal-to-noise ratio to better resolve the detection events.

With any of these techniques, the detection event signal after post-processing can be amplified and then discriminated by a comparator, creating a digital output signal from the detector.

\section{The semiconductor-based self-differencing detector}

\begin{figure}[b]
\centering
\includegraphics[width=12cm]{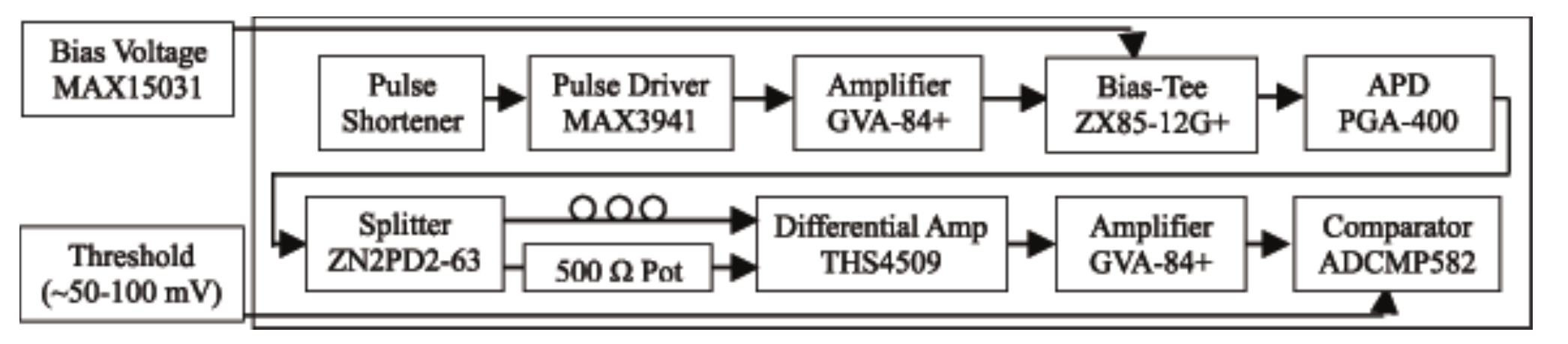}
  \caption{A block diagram of the electrical components of the detector as described in the text.  The top row (bottom row) shows the components for the gate signal generation (post-processing). }
\label{APDschematic}
\end{figure}

As shown in Fig.~\ref{apdFunctionalDiagram} and \ref{APDschematic}, our detector consists of several components, each described below.  A Princeton Lightwave PGA-400 Avalanche Photodiode in series with a 50~$\Omega$ resistor is mounted inside a radio-frequency shielding aluminum box, and cooled with a three-stage thermo-electric cooler (TEC), 5~cm heatsink and fan.  The temperature of the APD is regulated with a TEC-Microsystems DX5100 Temperature Controller and a MOSFET current amplifier to $(30.0\pm0.01)^{o}$C.

A 200 MHz LVPECL signal from a reference is pulse-shortened to create 900 ps full-width half-maximum (FWHM) pulses that are coupled through a capacitor to trigger a Maxim MAX3942 10 Gbps Modulator Driver.  These pulses are made single-ended with a Minicircuits ADT1-1WT Transformer, and amplified with a Minicircuits GVA-84+ Amplifier Evaluation Board.  The DC bias voltage is generated with a Maxim MAX15031 APD Bias Power Supply Evaluation Kit, and is combined with the gating pulse on a Minicircuits ZX85-12G Bias-Tee and sent to the APD cathode.  The detector was designed to operate at 200~MHz, with a duty cycle of approximately 20\%, to meet the requirements of our quantum key distribution (QKD) system~\cite{ItzelinPreparation}; yet, it can be gated at 500~MHz ($\sim50\%$ duty cycle) without changing the gate pulse.

The output of the APD is split using a Minicircuits ZN2PD2-63-S+ Power Splitter, and one path is delayed in approximately 109~cm of RG316 coaxial cable.  The other path is attenuated with a Bourns 500~$\Omega$ potentiometer to match the loss in the delay line, and then the signals are subtracted with a Texas Instruments THS4509 Differential Amplifier Evaluation Board.  The THS4509 is a wideband, fully-differential operational amplifier with a small-signal bandwidth (at 10~dB gain) of 1.9~GHz (an equivalent input risetime of approximately 250~ps).  The non-inverting output is amplified by another GVA-84+ amplifier and sent to the non-inverting input of an Analog ADCMP582 High-speed Comparator Evaluation Board, and the inverting output is used as a monitor signal.  After thresholding the signal at $\sim$80 mV, the differential output of the comparator is an LVPECL digital pulse with an average FWHM of 217~ps.  To enable interfacing with subsequent data collection, the output pulse is lengthened to 5 ns and then translated into a LVTTL signal.

\subsection{Characterizing the self-differencing circuit}

\begin{figure}[b]
\centering
\includegraphics[width=13cm]{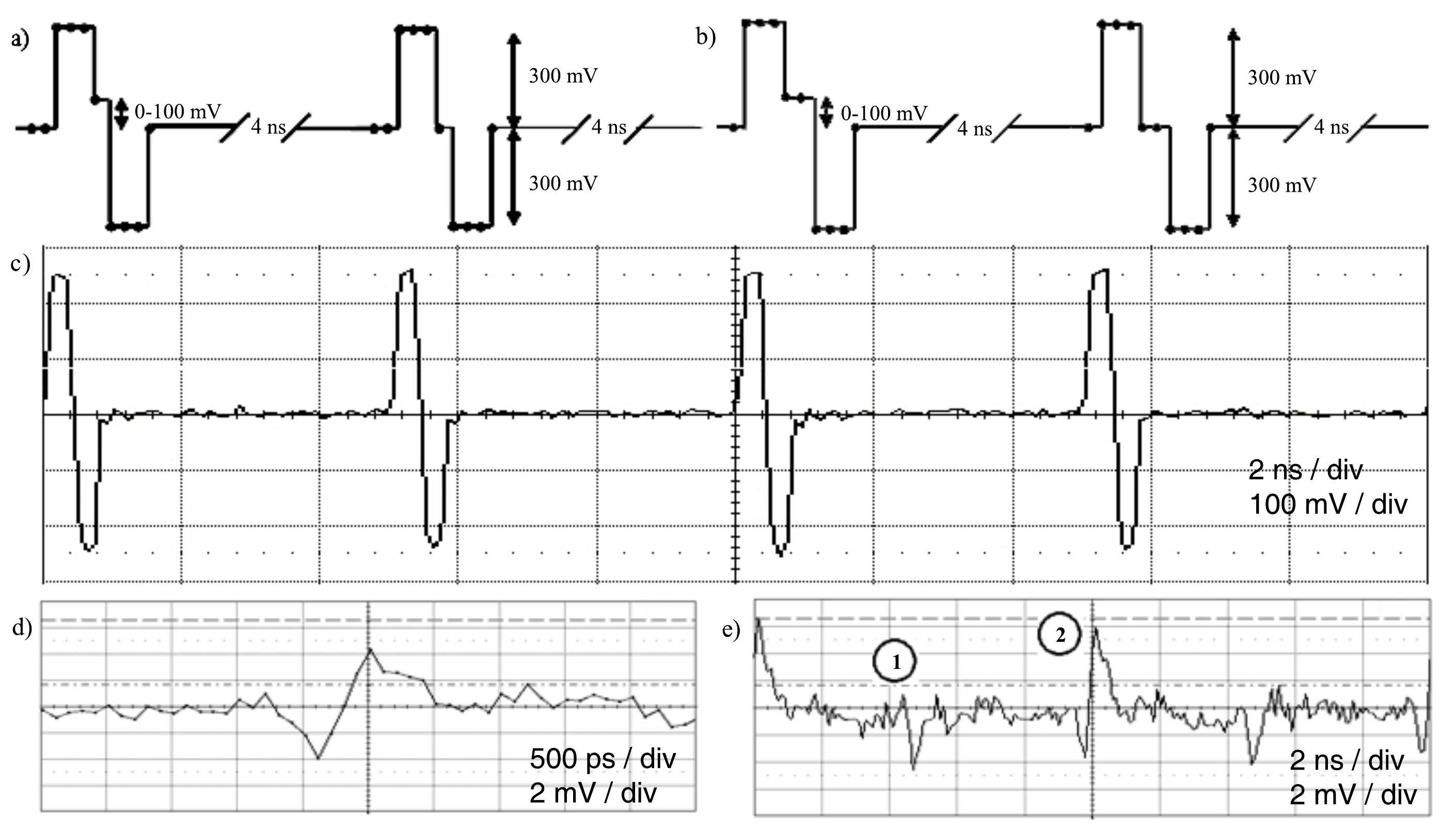}
  \caption{a)  The APD output simulation signal schema for 100~ps pseudo-detection events b) for 200~ps pseudo-detection events.  c) A 6~GHz oscilloscope trace of the simulated APD capacitive response generated by the AWG with a 100~ps, 100~mV detection event on every other pulse.  d) The remnants of the test signal on one gate after the differential amplifier (with 10~dB gain of the amplifier) with no detection events.  e) Output of the differential amplifier with 100~ps, 100~mV pseudo-detection events on every other gate.  The maximum positive voltages on gates 1 and 2 were measured for the characterization of the circuit shown in Figure~\ref{diffAmpDetectionPlots}.}
\label{APDsimulationSchematicFull}
\end{figure}

To allow for a detailed characterization of the self-differencing circuit, we used simulated signals generated by a Tektronix AWG7102 10~GSamples/s Arbitrary Waveform Generator (AWG) instead of actual APD outputs. These \textit{test signals} match the output of the APD as closely as possible as we found the performance of the subtraction circuit to depend strongly on the shape of the input signal. As shown in Fig. \ref{APDsimulationSchematicFull}, we achieve over 25~dB suppression of the simulated capacitive response. In contrast, if, after the circuit is tuned, a 200~MHz clock signal with similar amplitude is sent through the circuit, the suppression is only $\sim$14~dB, due to the differing frequency components.

\begin{figure}[t]
\centering
\includegraphics[width=13cm]{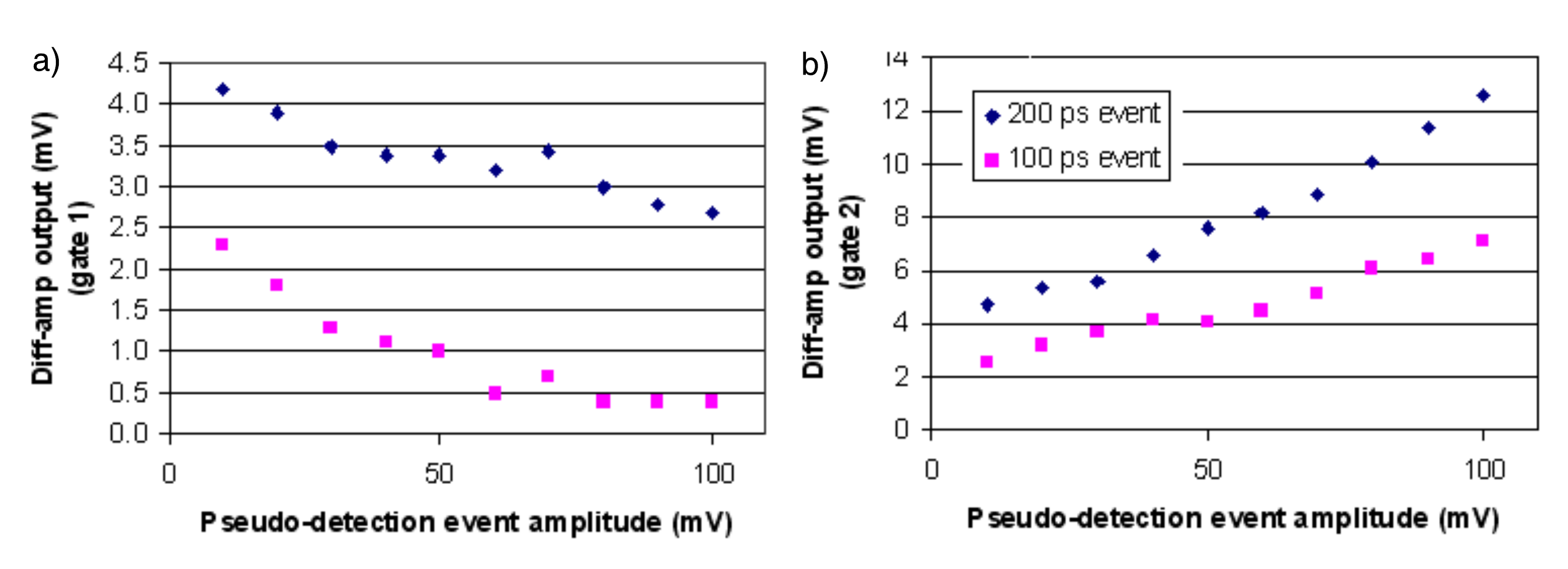}
  \caption{The maximum positive voltage of signals in two successive gates, output from the differential amplifier (shown in Fig.~\ref{APDsimulationSchematicFull}~(e)) in response to 100 and 200~ps pseudo-detection events of 10-100~mV. The left-hand figure shows results from gate 1, the right-hand figure from gate 2.}
\label{diffAmpDetectionPlots}
\end{figure}

The test signal programmed to simulate two APD output periods is shown in Fig.~\ref{APDsimulationSchematicFull}(a) and (b), one period with and one without a pseudo-detection event, at a repetition rate of 100 MHz. The capacitive response is modeled by three voltage samples, each one lasting 100~ps, of +300~mV, followed by a sample or two with zero voltage and three samples of  -300~mV. The intermediate samples in one gate were used to mimic detection events of 100 or 200~ps duration with voltage levels varying between 10 and 100 mV. The generated test signal across four gate periods is shown in Fig.~\ref{APDsimulationSchematicFull}(c).  It was sent through the splitter, and the coaxial cable delay-line was then cut and soldered to achieve the maximum suppression of the simulated capacitive response (with no detection event signal) on the output of the differential amplifier as shown in Fig.~\ref{APDsimulationSchematicFull}(d).  We then adjusted the pseudo-detection event signal from 10-100 mV, and observed detection events on every other gate as expected. As the detection event voltage depends on the number of photons detected per light pulse~\cite{ShieldsNumberAPD}, it is useful to know the relation between the amplitudes of a detection event and the differential amplifier output. As such, the maximum positive voltages of two successive gates, shown in Fig.~\ref{APDsimulationSchematicFull}(e), were measured from the non-inverting output of the differential amplifier; the results are plotted in Fig.~\ref{diffAmpDetectionPlots}. To discriminate the pulses with the comparator, and given the gain in the post-processing, we need the amplitude of a pseudo-detection event to exceed 100~mV (50mV) for events of 100~ps (200~ps) duration.

\subsection{Testing the detector}

\begin{figure}[t]
\centering
\includegraphics[width=11cm]{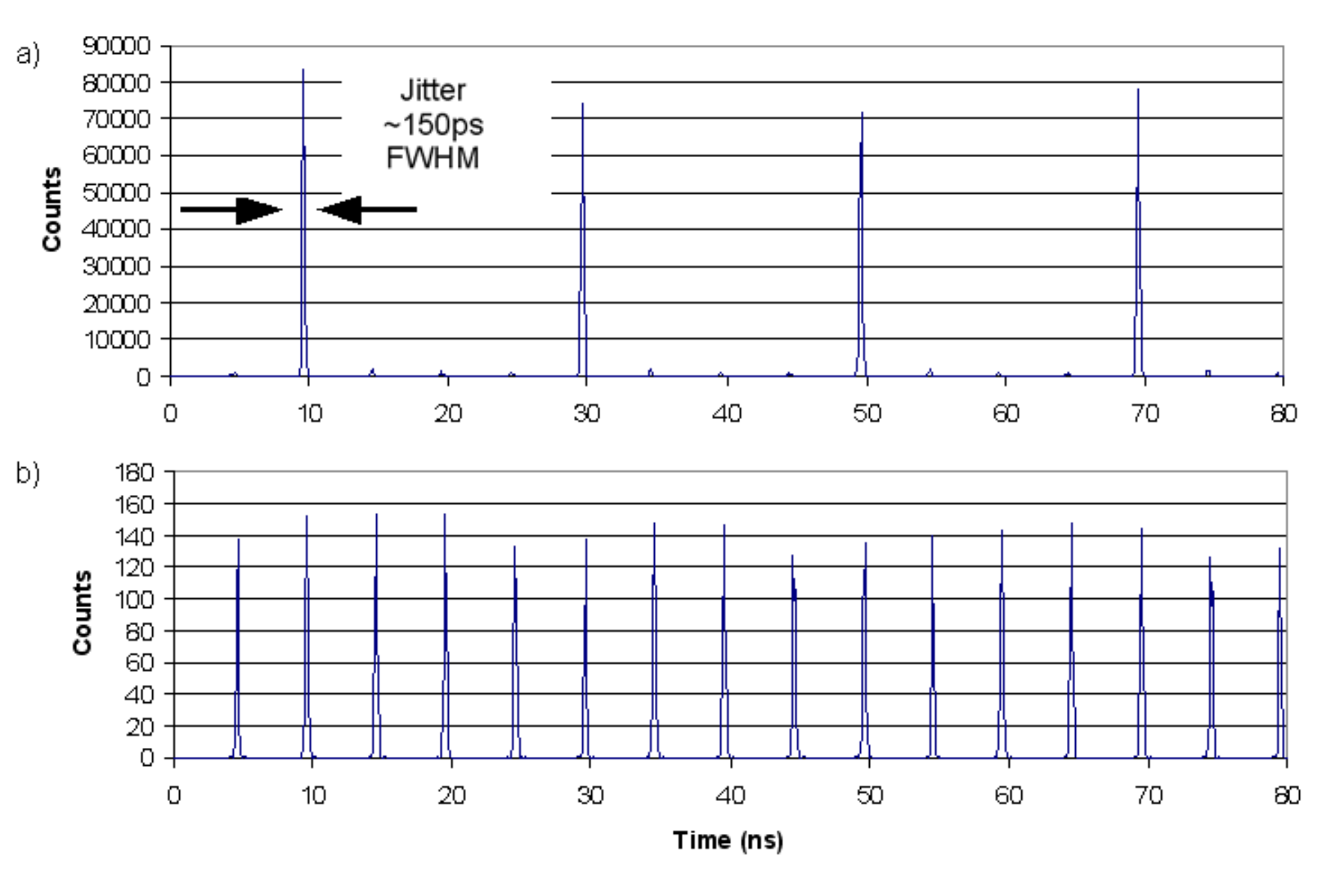}
  \caption{a) A TDC plot showing the single-photon detection capability of the detector.  With 50~MHz laser pulses incident on the detector, counts are seen to build up in gates separated by 20~ns. The FWHM of the detection peaks, i.e. the detection time jitter, is $\sim$ 150~ps. b) Same measurement as before, however without laser illumination. The dark counts are random and uniformly distributed in each gate.}
\label{detectorCharacterizations}
\end{figure}

\begin{figure}[b]
\centering
\includegraphics[width=10cm]{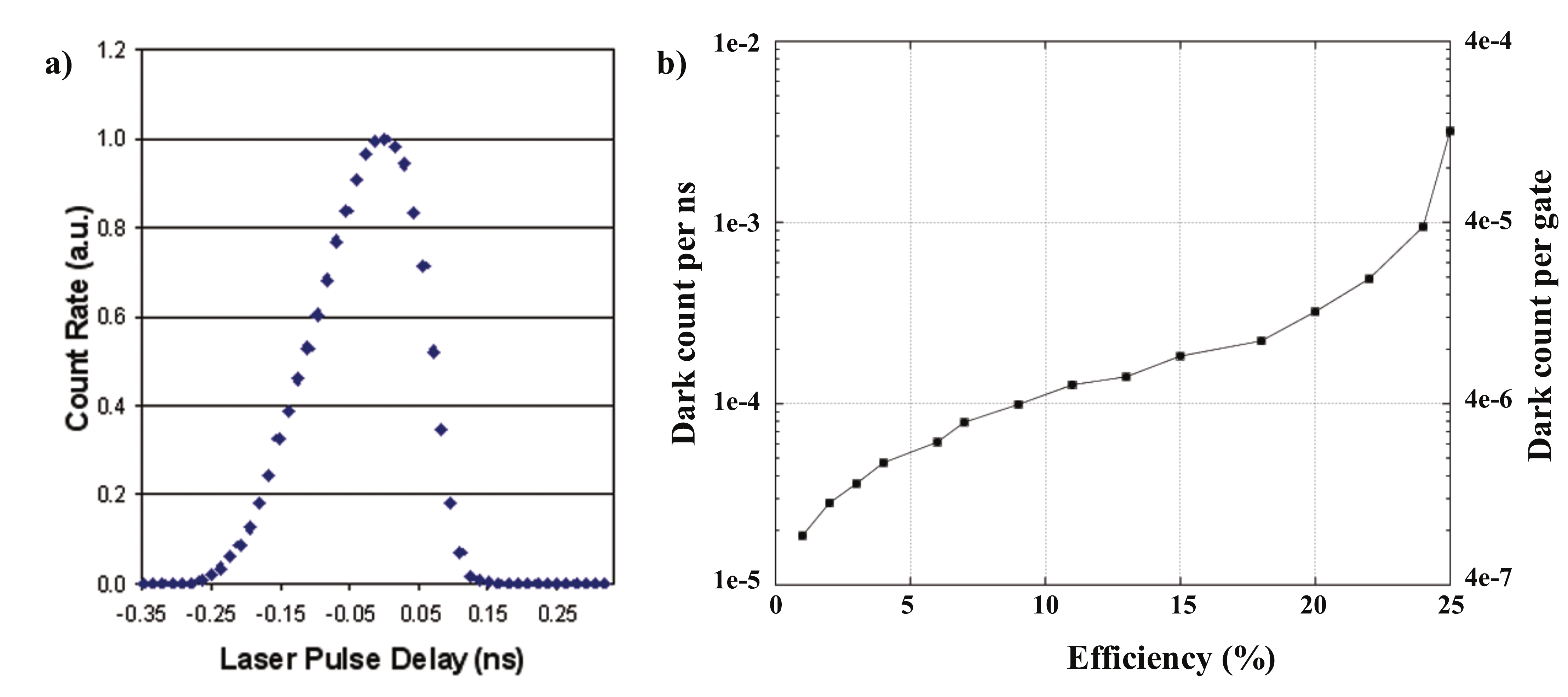}
  \caption{a) The photon count rate as a function of the laser pulse delay.  b) The dark count probability per ns, $P_{dc}^{ns}$, and per gate, $P_{dc}$, as a function of the detection efficiency $\eta$.}
\label{detectorCharacterizations1}
\end{figure}

To test the detector, we connect the now tuned self-differencing circuit to the APD. Laser pulses of 500~ps FWHM at 1550 nm wavelength, attenuated to an average of one photon-per-pulse, were coupled into the APD at a 50~MHz rate, and synchronized with the 200~MHz detector gates.  The output of the detector was used to stop a time-to-digital converter (TDC); it was triggered by a synchronized 1 MHz signal and recorded arrival times of detector output signals with respect to the trigger. Fig.~\ref{detectorCharacterizations}(a) is a plot of the output of the TDC showing that photons were counted in gates every 20 ns, as expected, with few dark counts in between the illuminated gates. From the detection events in the illuminated gates, we determine the detection time jitter to be $\sim$150~ps FWHM.  Figure~\ref{detectorCharacterizations}(b) shows that dark counts are uniformly distributed amongst all gates (the small undulation is due to aliasing), and that there are no counts in-between.

Next, we scan the laser pulse across the detector gate to measure the active time $\Delta t$ of the detector (this measurement was done with an average photon number of 0.01 per pulse).  As previously reported \cite{InoueSineWaveGating,ShieldsHighSpeedSPDNearInfrared,GenevaPracticalFastGateRate}, Fig.~\ref{detectorCharacterizations1}(a) shows that it is shorter than the duration of the electrical gate. For our detector, we find the active time to be $\Delta t=$200~ps (FWHM).  As expected, the count rate falls to zero outside of the active time. Hence, the duty cycle of the detector is 4\%; if gated at 500~MHz, we expect it to increase to 10\%.

The same setup also allows us to measure the detection efficiency $\eta$. It is calculated by~\cite{GenevaPracticalFastGateRate}:

\begin{equation}
\eta = (1/\mu)ln((1-R_{dc}/f_{g})/(1-R_{pd}/f_{p}))
\end{equation}

\noindent where $\mu$ is the average photon number, $R_{dc}$ is the dark count rate, $R_{pd}$ is the photon detection rate (i.e. the count rate in illuminated gates), and $f_{g}$ and $f_{p}$ are the gating and laser pulse frequencies, respectively.  The dark count probability per nanosecond $P_{dc}^{ns}=P_{dc}/(f_{g}\Delta t)$, as well as the dark count per gate probability $P_{dc}$ are plotted as a function of the efficiency in Fig.~\ref{detectorCharacterizations1}(b).  At $10\%$ efficiency, the detector has a dark count per gate probability of 4x10$^{-6}$, corresponding to a dark count probability of 1x10$^{-4}$/ns.  Note that no deadtime, or ``count-off time" is applied to the circuit in response to a detection event.

\begin{figure}[t]
\centering
\includegraphics[width=9cm]{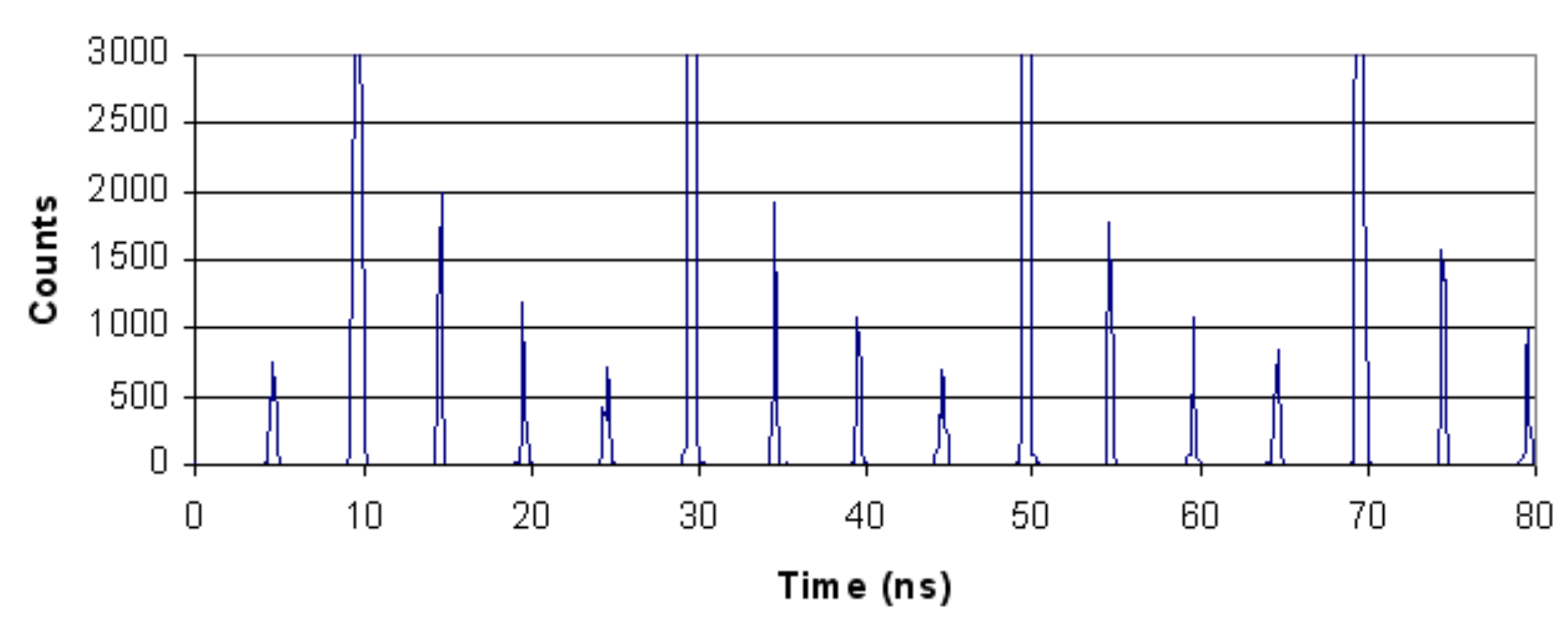}
  \caption{A zoom-in of the non-illuminated gates in Fig.~\ref{detectorCharacterizations} shows the expected decreasing afterpulse probability. }
\label{afterpulsingJitter}
\end{figure}

Figure~\ref{afterpulsingJitter}(a) shows a zoom-in of the non-illuminated gates in Fig.~\ref{detectorCharacterizations}(a), showing a decay of afterpulsing events in time after detection events.  The  afterpulsing probability per detection event is calculated as  \cite{InoueSineWaveGating,ShieldsHighSpeedSPDNearInfrared,BenMichaelPLIafterpulsing}:

\begin{equation}
P_{a} = \frac{(C_{ni}-C_{dc})\cdot R}{(C_{i}-C_{ni})}.
\end{equation}

\noindent Here, $C_{ni}$ is the average number of counts per non-illuminated gate, $C_{i}$ the average number of counts per illuminated gate, $C_{dc}$ the number of dark counts per gate (see Fig.~\ref{detectorCharacterizations}), and  $R=f_{g}/f_{p}$. At an efficiency of $6.4\%$, we find $P_{a}=6.3\%$.  If we normalize to the active time~\cite{GenevaPracticalFastGateRate}:

\begin{equation}
P_{a}^{ns} \sim \frac{P_{a}\cdot f_{P}\mu\eta}{f_{g}\Delta t}
\end{equation}

\noindent we have an afterpulsing probability per nanosecond $P_{a}^{ns}=5\times10^{-3}$/ns.  Note as well that $P_{a}$ and $P_{a}^{ns}$ include afterpulses from dark counts and other afterpulses.

\begin{figure}[b]
\centering
\includegraphics[width=10cm]{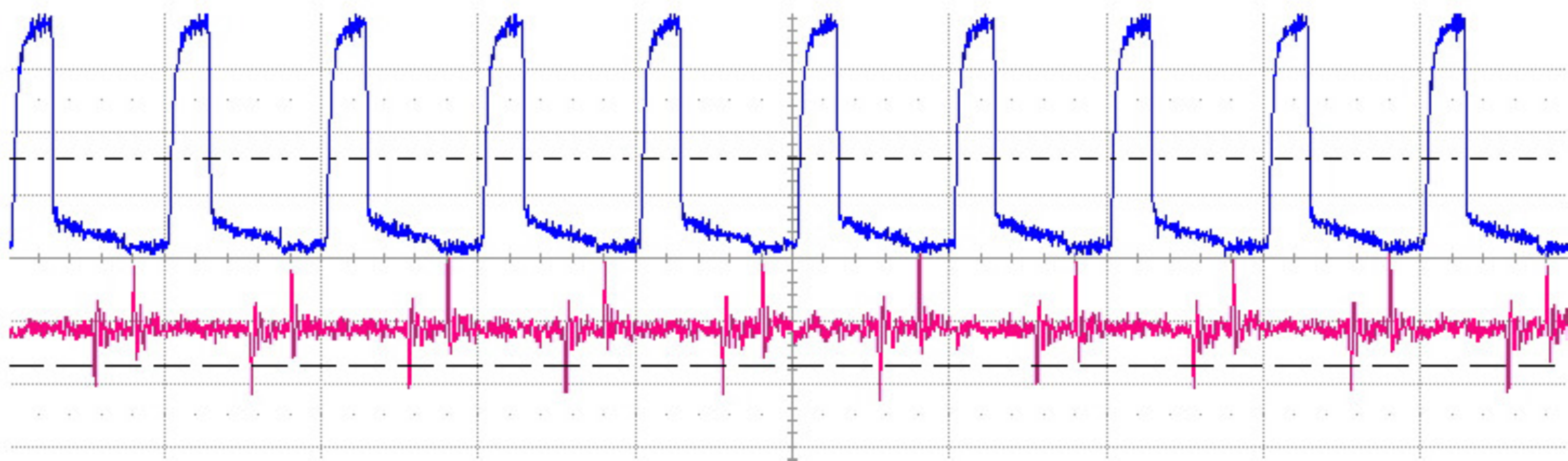}
  \caption{Oscilloscope trace showing signals generated from detecting strong laser pulses constituting quantum frames in our QKD system; the bottom curve shows the output of the differential amplifier, and the top curve depicts the detector output. The scales are 20~ns per division and 500~mV (10~mV) for the top (bottom) trace.}
\label{classicalHeaderDetection}
\end{figure}

\section{Applications}

Single-photon detectors gated at high rates are important for a large variety of experiments both in the field of quantum optics as well as in quantum communication. For instance, such a detector allows an efficient characterization of a source of non-classical light by means of measurements of the cross-correlation coefficient~\cite{ClauserCrossCorrelation,ChrisDinPreparation}, where measurement times are determined by detector gate rates. As another example, we have used the detector in our quantum key distribution (QKD) system\cite{ItzelQC2QKD,ItzelinPreparation}. The system alternates between sequences of attenuated laser pulses encoding quantum bits, and strong pulses of light forming classical control frames. The latter include information that enables routing of quantum information to different users, clock synchronization, and assessing and compensating of time-varying birefringence in the quantum channel. The detector serves multiple purposes. First, it allows increasing the quantum bit generation rate to 100 MHz, thereby removing a major bottleneck towards high secret key rates. Second, if the gating pulse of the detector is switched off, the APD operates in the standard (linear) photodiode mode, and can detect the 500~ps classical pulses forming the control frames; Fig.~\ref{classicalHeaderDetection} shows the differential amplifier and detector outputs to an input bit-string of '1's.  Furthermore, we point out that one can also monitor the current of the bias voltage source. In addition to detecting information about the control frames, it may allow detection of fake-state attacks~\cite{FakedStateMakarov,FakedStateShields}.

\section{Discussion and conclusion}

 In the current design, comprising inexpensive and commercially available components, our detector performs with specifications suitable for experiments requiring high-rate single-photon detection.  The versatility of the detector can be further improved (and the detector made tunable to a variable gate frequency) if an adjustable delay line is included in the self-differencing circuit~\cite{ShieldsMultiGHzSD}. Furthermore, we note that the delay line, which is currently realized using a coaxial cable, can be implemented onto a printed circuit board~\cite{DelayLineMeander,DelayLineGenetic}, thereby allowing for additional integration. However, the $\sim$3 dB loss calculated \cite{RogersMIC} for a 5 ns delay restricts the use of PCB-based delays to shorter times. As an example, a 2 ns delay (increasing the gate-rate to 500 MHz), implemented by means of a 6.235" long trace ($1"=2.54$~cm) on a 0.06" thick RO4350 PCB substrate with 1/2 ounce copper (with trace width of 0.14579" for 50~$\Omega$ impedance matching), features 1.16 dB attenuation for a 10 GHz-bandwidth signal. This is comparable to the 1.31 dB attenuation that we measured for our current delay, implemented using the 109~cm-long RG316 cable. We point out that increasing the gate rate to and beyond 1 GHz, which is associated with increased signal bandwidth, will eventually face limitations due to the limited bandwidth of the electronic components used in this design. For instance, the differential amplifier features only 1.9~GHz bandwidth, and the gain of the GVA-84+ amplifier is restricted to signals below $\sim$7 GHz. To give an example how performance can be improved by using better components, we note that when a state-of-the-art 25~GHz SHF amplifier was used instead of the GVA-84+ amplifier, the ratio of efficiency to dark counts was doubled.

In conclusion, we have characterized a 200~MHz gate-rate single-photon detector that uses an InGaAs/InP avalanche photodiode and a self-differencing post-processing circuit implemented with a differential amplifier. The self-differencing circuit was tuned with a test signal that simulated the capacitive response of the APD, and shows promise to be fully integrated on a printed circuit board. The detector can be operated up to a 500~MHz gate rate, and shows typical characteristics for a high-rate single-photon detector. Furthermore, it can be operated in the standard (linear) photodiode regime, which benefits applications that require detecting single photons as well as strong light signals, as exemplified with our quantum key distribution system.



\section*{Acknowledgments}

The authors gratefully acknowledge Vladimir~Kiselyov for electronics support, and Steve~Hosier, Joshua~Slater, Philip~Chan and F\'{e}lix Bussi\`{e}res for many insightful discussions.  This work is supported by General Dynamics Canada, Alberta's Informatics Circle of Research Excellence (iCORE, now a part of Alberta Innovates), the National Science and Engineering Research Council of Canada (NSERC), QuantumWorks, the Canada Foundation for Innovation (CFI), Alberta Advanced Education and Technology (AET), and the Mexican Consejo Nacional de Ciencia y Tecnolog\'ia (CONACYT).

\end{document}